\documentclass[pre,superscriptaddress,twocolumn,footinbib]{revtex4-1}
\pdfoutput=1
\usepackage[colorlinks=true,urlcolor=blue]{hyperref}
\usepackage[separate-uncertainty=true]{siunitx} 
\usepackage{chemmacros}
\usepackage{amsfonts}
\usepackage{graphicx}
\usepackage{placeins}
\usepackage{amsmath}
\usepackage{gensymb}
\usepackage{xcolor}
\usepackage{color}
\usepackage{bm}

\DeclareSIUnit\Molar{\textsc{m}}

\definecolor{red}{rgb}{0.75,0,0}
\definecolor{blue}{rgb}{0,0,0.75}
\definecolor{green}{rgb}{0,0.5,0}

\begin{document}

\title{Lipid exchange enhances geometric pinning\\ in multicomponent membranes on patterned substrates}

\author{Melissa Rinaldin}
\affiliation{Instituut-Lorentz, Universiteit Leiden, P.O. Box 9506, 2300 RA Leiden, The Netherlands}
\affiliation{Huygens-Kamerlingh Onnes Lab, Universiteit Leiden, P. O. Box 9504, 2300 RA Leiden, The Netherlands}
\author{Piermarco Fonda}
\affiliation{Instituut-Lorentz, Universiteit Leiden, P.O. Box 9506, 2300 RA Leiden, The Netherlands}
\author{Luca Giomi}
\email{giomi@lorentz.leidenuniv.nl}
\affiliation{Instituut-Lorentz, Universiteit Leiden, P.O. Box 9506, 2300 RA Leiden, The Netherlands}
\author{Daniela J. Kraft}
\email{kraft@physics.leidenuniv.nl}
\affiliation{Huygens-Kamerlingh Onnes Lab, Universiteit Leiden, P. O. Box 9504, 2300 RA Leiden, The Netherlands}

\date{\today}

\begin{abstract}
Experiments on supported lipid bilayers featuring liquid ordered/disordered domains have shown that the spatial arrangement of the lipid domains and their chemical composition are strongly affected by the curvature of the substrate. Furthermore, theoretical predictions suggest that both these effects are intimately related with the closed topology of the bilayer. In this work, we test this hypothesis by fabricating supported membranes consisting of colloidal particles of various shapes lying on a flat substrate. A single lipid bilayer coats both colloids and substrate, allowing local lipid exchange between them, thus rendering the system thermodynamically open, i.e. able to exchange heat and molecules with an external reservoir in the neighborhood of the colloid. By reconstructing the Gibbs phase diagram for this system, we demonstrate that the free-energy landscape is directly influenced by the geometry of the colloid. In addition, we find that local lipid exchange enhances the pinning of the liquid disordered phase in highly curved regions. This allows us to provide estimates of the bending moduli difference of the domains. Finally, by combining experimental and numerical data, we forecast the outcome of possible experiments on catenoidal and conical necks and show that these geometries could greatly improve the precision of the current estimates of the bending moduli.
\end{abstract}

\maketitle

\section{Introduction}

Multicomponent artificial lipid membranes, consisting of a ternary mixture of cholesterol and phospholipids, undergo liquid-liquid phase separation at specific temperature and lipid chemical composition~\cite{veatch2002organization,veatch2003separation,levental2009cholesterol, wesolowska2009giant}. When they assemble into a bilayer, at high temperatures, the lipids form a uniformly mixed membrane, while at lower temperatures, they spontaneously separate into two different liquid phases, known as liquid ordered (LO) and liquid disordered (LD). The LO phase is rich in saturated lipids and cholesterol while the LD phase is rich in unsaturated lipids \cite{mukherjee2004membrane, ayuyan2008raft, almeida2003sphingomyelin, baumgart2007fluorescence, sezgin2012partitioning}. 

The LO phase is more packed and thicker than the LD phase, implying that LO domains are less prone to bending and splay deformations. These domains are preferentially localised in regions of relatively low mean curvature and avoid regions of high negative Gaussian curvature. Experimental evidence of this phenomenon has been obtained in experiments with giant unilamellar vesicles (GUVs) \cite{baumgart2003imaging,baumgart2005membrane,hess2007shape,semrau2009membrane,heinrich2010dynamic,sorre2009curvature}, with supported lipid bilayers (SLBs) on patterned substrates \cite{Parthasarathy2006,subramaniam2010particle, feriani2015soft} and, most recently, on scaffolded lipid vesicles (SLVs), i.e. membrane-coated colloidal particles \cite{Rinaldin2018}. While GUVs can change shape during the phase separation process, SLBs and SLVs are scaffolded. Therefore, they can mimic the supporting property of the actin cytoskeleton in cells and allow for the fabrication of stable bilayers for atomic force microscopy (AFM) studies \cite{miller2018substrate, goodchild2019substrate}, for applications in bio-sensing \cite{Chemburu2010}, in cell biology \cite{Madwar2015}, and in drug delivery \cite{ashley2012delivery}. 

Experiments with GUVs have shown that there is a correlation between positioning of soft domains and bilayer curvature. However, the large variety of possible shapes and the fact that vesicles change their shape dynamically while phase-separating has prevented a quantitative characterisation of this phenomenon, owing to the scarce reproducibility. Conversely, experiments with SLBs on corrugated glass \cite{Parthasarathy2006} and flat substrates with hemispherical asperities \cite{subramaniam2010particle, feriani2015soft} have shown that the LD domains are consistently geometrically pinned to regions of high mean curvature. This pinning was also observed in experiments with SLVs of symmetric and asymmetric dumbbells \cite{Rinaldin2018} albeit only for LD domains relatively small compared to the total vesicle size. Even more remarkably, the most common configurations observed in anisotropic SLVs, featured {\em antimixing}: i.e. a state in which the lipids are mixed and yet strongly partitioned on regions with different curvature \cite{Rinaldin2018,fonda2019thermodynamic}. 
Further numerical studies \cite{fonda2019thermodynamic,fonda2018interface} have shown that such a behaviour depends crucially on whether the membrane is open or closed, i.e. on whether the system can exchange lipids with an external reservoir. In fact, open membranes are expected to expel saturated lipids from regions of high curvature into the reservoir. In the present work, we test this hypothesis by studying supported lipid bilayers on substrates patterned with colloidal particles. In this system, there is a continuous exchange of lipids between the coated colloid and the environment. Consistent with the theoretical predictions \cite{fonda2019thermodynamic}, we indeed observe that the lipid exchange enhances geometric pinning (i.e. the correlation between phase domain location and curvature), but suppresses antimixing and, in general, any form of non-homogeneous mixing of the lipids.

\begin{figure}
\includegraphics[width=1\linewidth]{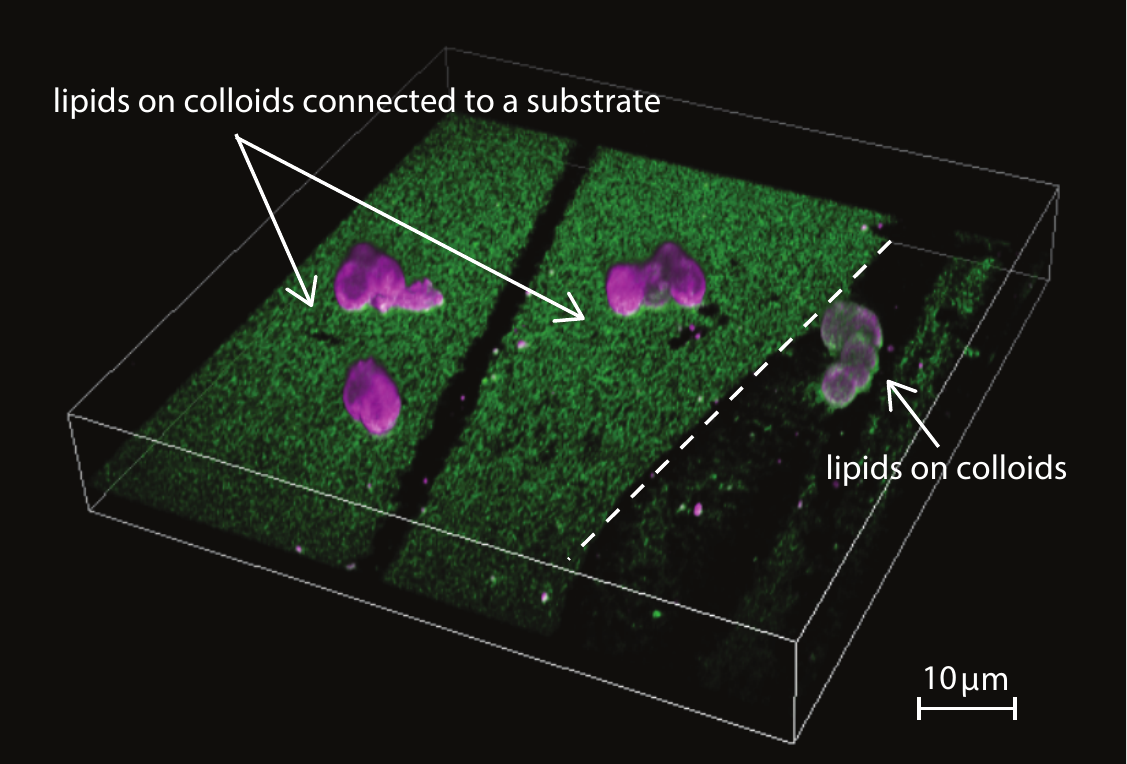}
\caption{Fluorescence microscopy image showing different coatings of colloidal particles on a substrate. On the left, a single lipid membrane covers both substrate and colloidal particles. Green and magenta colours indicate the LO and LD phase, respectively. On the right, the lipid bilayer covers only two isolated colloids. At the same lipid composition, the left membrane is phase-separated, with the LD phases pinned onto the colloids, while the right membrane is in the mixed state.}
\label{fig:introfig}
\end{figure}

In Fig. \ref{fig:introfig}, we show an example of our experimental results.  On the left side, the lipid bilayer covers the colloids and part of the substrate. The bilayer on the planar substrate is enriched in saturated lipids, shown in green, while the bilayer on the colloidal particles is richer in unsaturated lipids, labeled in magenta. On the right part of the picture, the bilayer is present only on the colloid and not the substrate. Its color is neither pure green nor pure magenta, indicating that the different lipids are well mixed. Since the overall lipid composition of the mixture is the same, Fig. \ref{fig:introfig} shows that the lipid exchange determines whether phase separation occurs or not and, if it occurs, that it leads to a geometrically pinned configuration.

In this paper, we analyse the effect of local lipid exchange and compare our results with our previous work on SLVs \cite{Rinaldin2018}. We emphasize that, while at the scale of a single colloid there is exchange of lipids with the substrate, at the scale of the entire system the total amount of lipids is fixed. This allows us to sample the Gibbs phase triangle of the lipid mixture by varying the total membrane composition, while rendering the colloid a thermodynamically open system, that is able to exchange lipids with the substrate, which, in turn, acts as a reservoir. 
Furthermore by comparing experimental and numerical data, we are able to estimate the material properties of the membrane and show that these are consistent with previous measurements made on multicomponent GUVs of similar compositions. Finally, we forecast the outcome of possible experiments on catenoidal and conical necks and show that these geometries could greatly improve the precision of the current estimates of the bending moduli.

\section{Experimental setup}

\subsection{Reagents}

The lipids 1-palmitoyl-2-oleoyl-\emph{sn}-glycero-3-phosphocholine (POPC), porcine brain sphingomyelin (BSM), ovine wool cholesterol (chol), 1,2-dioleoyl-\emph{sn}-glycero-3-phosphoethanolamine-N-lissamine rhodamine
B sulfonyl 18:1 (Liss Rhod PE), N-[11-(dipyrrometheneboron difluoride)undecanoyl]-Derythro-sphingosylphosphorylcholine, and \iupac{1,2-di|o|le|oyl-sn-gly|ce|ro-3-phos|pho|e|tha|nol|a|mine-N-[me|tho|xy|(po|ly|e|thy|lene|gly|col)|-2000]} (DOPE-PEG(2000)), were purchased by Avanti Polar Lipids and stored at -20\celsius.

\subsection{Substrates and colloids}

Borosite glass coverslips 1 mm thick were purchased by VWR international. Silica spheres were purchased from Microparticles GmbH (\SIlist{2.06\pm0.05;7.00\pm0.29}{\micro\meter} in diameter) and Fluka (\SIlist{3.00\pm0.25}{\micro\meter} in diameter). Polystyrene-3-(Trimethoxysilyl)propyl methacrylate (PS-TPM) isotropic dumbbell particles with long axis of \SIlist{5.23\pm0.05}{\micro\meter} and ratio of the diameters of the two lobes of \SIlist{0.98\pm0.04} and asymmetric dumbbell-shaped particles with long axis of \SIlist{4.01\pm0.04}{\micro\meter} and ratio of the diameters of the two lobes of \SIlist{0.57\pm0.02}  were synthesized by making a protrusion from swollen PS-TPM particles \cite{kim2006synthesis} and coated with silica \cite{wang2013shape}. Hematite cubic particles were made following \cite{sugimoto1993formation}, coated with silica \cite{Rossi2011} and treated with HCl to remove the hematite core to obtain cubic shells with a resulting corner-to-corner distance of $1.76 \pm 0.08$ \si{\micro\meter} and m-value of $3.3 \pm 0.6$. Hepes buffer was made with 115 mM NaCl, 1.2 mM $\mathrm{CaCl}_2$, 1.2 mM $\mathrm{MgCl}_2$, 2.4 mM $\mathrm{K_2HPO_4}$ and 20 mM Hepes, all purchased from Sigma Aldrich. All chemicals were used as delivered.

\subsection{Sample preparation}

\subsubsection{Substrates} 

Particular attention was given to the cleaning of the substrates since it strongly influences the formation of the bilayer. Glass coverslips were cleaned while magnetic stirring in a solution of 0.2\% Hellmanex, ethanol and milliQ, and washed three times in milliQ after each step. The coverslips were kept in milliQ for a maximum of three days, in order to keep them hydrophilic.

All particles were washed three times in milliQ before being used in lipid coating experiments. To attach the colloids to a substrate, a dispersion of particles was then dried on the glass coverslip. Since the particles experience strong capillary forces during the drying process, they often form clusters. To suppress aggregation, we used a small quantity of particles, namely $\approx 50$ \si{\micro\liter} of 5 \si{\gram\liter^{-1}} colloid dispersion for each coverslip. For the experiments with cubic particles, we heated the sample to \si{500 \celsius} to burn the polymers used for the stabilisation of the cubic colloids from the surface, as shown in \cite{rinaldin2019colloid}. In some experiments with symmetric and asymmetric dumbbells and spheres, we also burnt the particles at 500 \si{\celsius} to remove the inner polymer network of polystyrene and TPM. We observed occasional breaking of the silica shells of calcinated silica-coated PS-TPM spheres and dumbbell-shaped colloids.

\begin{figure*}[ht]
\includegraphics[width=1\linewidth]{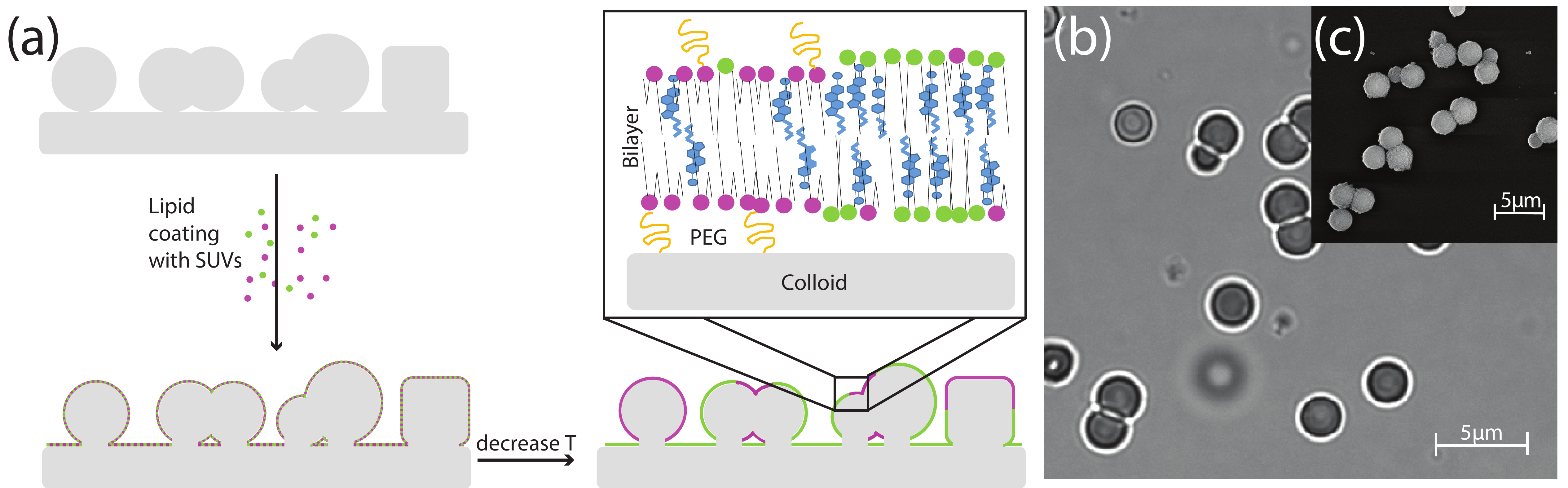}
\caption{
\textbf{(a)} Schematic overview of the experimental set-up. Colloids with a silica surface and different shapes are deposited onto a glass coverslip. At \SIlist{70}{\celsius}  the colloids and the substrate are coated with a lipid bilayer by deposition of Small Unilamellar Vesicles (SUVs). Upon lowering the temperature, the lipid bilayer undergoes phase separation. The schematic respresentations of lipids in the inset correspond to cholesterol (blue), SM (green), POPC (magenta) and PEG (yellow). The LO and LD phases are shown as green and magenta profiles, respectively.
\textbf{(b)} Bright field and \textbf{(c)} scanning electron microscopy image of a patterned substrate with symmetric and asymmetric dumbbell-shaped colloids.}
\label{fig:setup} 
\end{figure*} 
 
\subsubsection{Supported lipid bilayers}

A mixture of \SI{500}{\micro\gram} of POPC, BSM and Chol in different mole ratios with 0.2\% M/M DOPE Lissamine Rhodamine, 0.2\% M/M Topfluor Sphingomyelin and 5\% M/M DOPE-PEG in chloroform was prepared and the chloroform was evaporated in a vacuum chamber for 2 hours. The lipids were dispersed in Hepes buffer in 250 \si{\gram\liter^{-1}} concentration, and let to self-assemble into multilamellar vesicles during 30 minutes of vortexing. 
The dispersion was heated in an oven to 70\si{\celsius}, and then extruded 21 times with a mini-extruder (Avanti Polar Lipids) placed on a heating plate set at 70\si{\celsius} equipped with two 250 \si{\micro\liter} gas-tight syringes (Hamilton), four drain discs, and one nucleopore track-etch membrane (Whatman). Then, 50 \si{\micro\liter} of SUVs were added to a holder with the colloidal particles on the substrate and 1 \si{\milli\liter} of HEPES buffer and put in the oven at 70\si{\celsius} for one hour. After that, the sample was rinsed three times with HEPES buffer, in order to remove excess SUVs in dispersion. During each step described above, the SUVs were covered by aluminium foil to prevent bleaching of the dye and oxidation of unsaturated lipids. 

\subsection{Sample characterisation} 

The samples were imaged at room temperature with an inverted confocal microscope (Nikon Eclipse Ti-E) equipped with a Nikon A1R confocal scan head with Galvano and resonant scanning mirrors. A $100\times$ oil immersion objective (${\rm NA}=1.4$) was used for 488~\si{\nano\meter} and 561~\si{\nano\meter} lasers. These lasers were used to excite Top Fluor and Lissamine Rhodamine dyes, respectively. Lasers were passed through a quarter wave plate to avoid polarisation of the dyes and the emitted light was separated using 500-550~\si{\nano\meter} and 565-625~\si{\nano\meter} filters. The mobility of the bilayer on the colloids, on the substrates and in the contact region was checked by fluorescence recovery after photobleaching (FRAP) experiments (see Fig. 2 in the Supporting Information). 3D image stacks were acquired by scanning the sample in the $z-$direction with an MCL Nano-drive stage and reconstructed with Nikon AR software.

\section{Results and Discussion}

In previous work we showed how the organisation of LO and LD domains in SLVs is determined the interplay between their rigidity upon bending - which in turn results from the chemical composition of the LO and LD phases - and the curvature of the underlying substrate \cite{Rinaldin2018}. To study how these equilibrium configurations are affected by lipid exchange, we present here an experimental set-up in which the lipid composition is not conserved in the neighborhood of the colloid. To do so, we fabricate substrates patterned with colloidal particles and then coat them with a multicomponent lipid bilayer (see Fig. \ref{fig:setup}a). In this way, there is a continuous exchange of lipids between the portion of the membrane adhering to the colloid and the membrane lying on the flat substrate, which thus acts as a reservoir.  
We stress that, while there is lipid exchange between the portion of the membrane coating the colloid and the one lying on the substrate, the total lipid composition of the whole SLB is fixed and, as we will detail in the next section, its location in the Gibbs phase triangle determines the chances of phase separation occurring on the colloids.  

\subsection{The ternary phase diagram of a SLB patterned with colloids}

We studied the phase behavior of different SLBs by varying the concentration of POPC, BSM, and cholesterol on various substrates patterned with colloidal spheres, dumbbells and asymmetric dumbbells (referred to as ``snowman particles'' in the following). 
A homogeneous and mobile lipid coating is achieved by deposition of small unilamellar vesicles (SUVs) \cite{Richter2006}. The PEGylated phospholipid DOPE-PEG2000 was included in the mixture to increase the water layer between the surface of the particle and the colloid, hence improving the mobility of the bilayer on the rough silica surface. The fluorescent lipids DOPE-Rhodamine and TopFluor cholesterol were used to image the LD and the LO phases, respectively. In all images, these phases are shown in magenta (LD phase) and green (LO phase). The lipid coating was done above the critical transition temperature to keep both lipid bilayer and SUVs in the mixed state. Phase separation was induced by cooling the sample. The lipid mobility on the colloids, on the substrate and at the edge between the colloids and the substrate was confirmed by FRAP experiments, of which some examples are reported in the Supporting Information Fig. 2. 

\begin{figure*}
\includegraphics[width=1\linewidth]{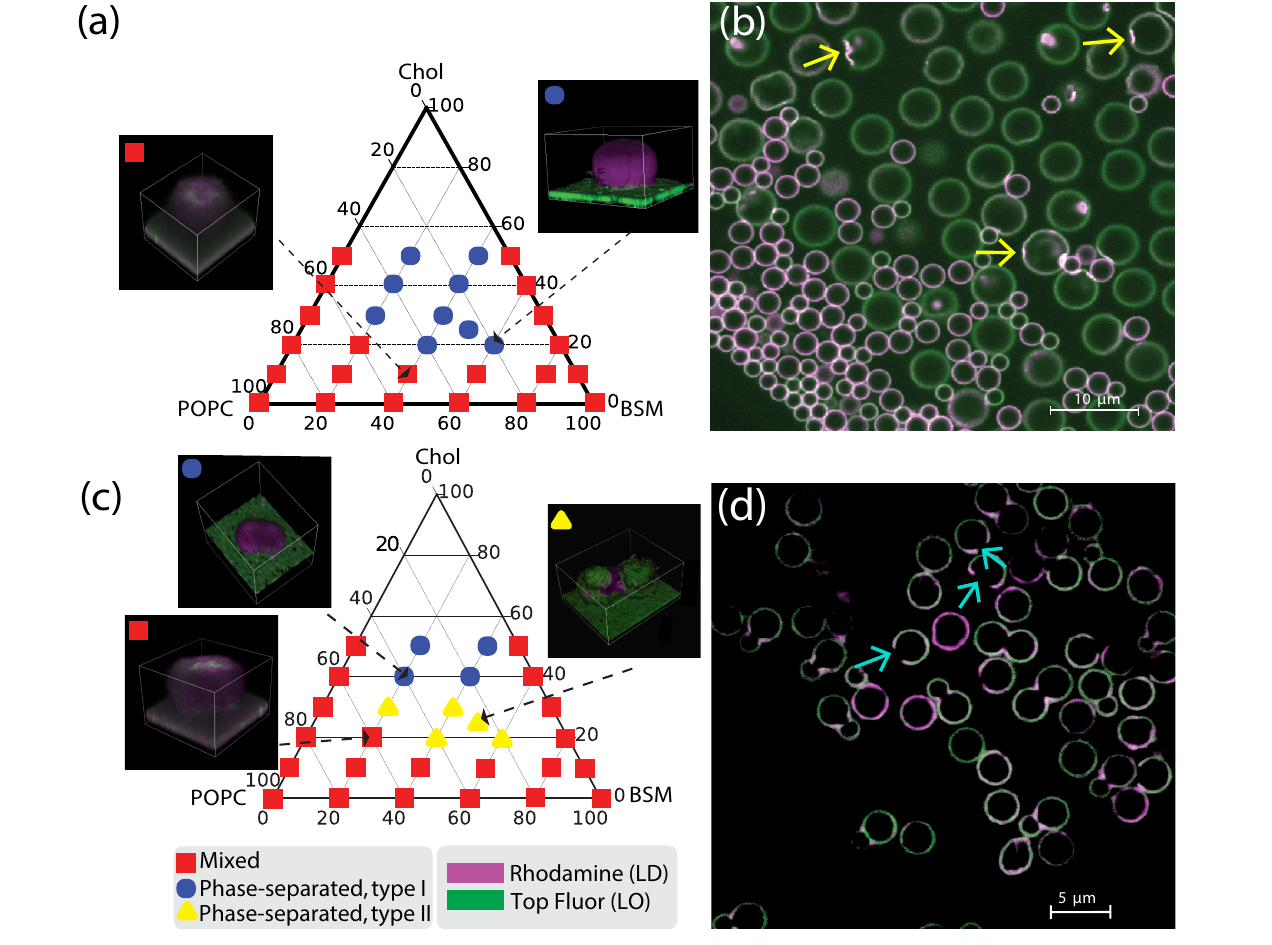}
\caption{
\textbf{(a)} Ternary phase diagram of a lipid bilayer on a substrate patterned with colloidal spheres. Depending on the overall composition, the membrane can be either in the mixed state (red squares) or in the phase separated state (blue circles). In the insets, we show 3D representative reconstructions of the bilayer in the mixed/phase separated states on a spherical substrate. LD and LO phases are represented in magenta and green, respectively. Single channel images of the insets are reported in the Supporting Information, Fig. 1.
\textbf{(b)} Equatorial view of a SLB with spherical colloidal particles of three different sizes. LD domains clearly prefer regions of higher curvature, i.e. spheres of smaller radii and dimples (indicated with yellow arrows).
\textbf{(c)} Ternary phase diagram of a lipid bilayer on a substrate patterned with dumbbell-shaped particles. 
In the insets we show 3D representative reconstructions of the bilayer in the mixed and phase separated state. 
\textbf{(d)} Equatorial view of a lipid bilayer supported on a plane with both symmetric and asymmetric dumbbell-shaped colloids. The LD phase is preferentially localized on the neck of the dumbbells. In this image the scaffold was first calcinated at \SI{450}{\celsius} so that only the colloidal silica shell survives, which in some cases is broken. For these particles, LD domains locate along broken edges, as shown by the light blue arrows.}
\label{fig:figure3} 
\end{figure*}

We show the effect of different colloidal shape on the stability phase diagram in Fig. \ref{fig:figure3}. Panel \ref{fig:figure3}a shows the Gibb phase triangle of a SLB on a substrate patterned with spherical colloids. In the lower center part of the ternary phase diagram, phase separation occurs, while at high concentrations of either of the three components the membrane remains in the mixed state. The insets show two representative 3D reconstructions of the samples in the mixed and the demixed state. In the reconstructions, the mixed phase appears white since the two fluorescent channels overlap. From these experiments, we were able to estimate the binodal line separating mixed from demixed states, obtaining similar results to experiments with GUVs with the same \cite{petruzielo2013phase} or similar \cite{veatch2003separation} lipid mixtures. We note that there are no data points for mixtures with more than 50\% cholesterol concentration, because such a region does not allow liquid-liquid coexistence on vesicles \cite{petruzielo2013phase}. In contrast to spherical SLVs, which feature varying area fractions of LO and LD phase depending on the composition,\cite{Rinaldin2018} spherical membranes in contact with a reservoir are either fully covered by an LD domain or in the mixed state. The interface of the demixed state is located at the contact line between the spheres and the substrate. 

Next, we studied how the size of the colloidal spheres affects the performance of geometric pinning. To this end, we prepared a patterned substrate consisting of multiple spheres of different radii, namely \SIlist{2.06\pm0.05;3.00\pm0.25; 7\pm0.29}{\micro\meter}. Interestingly, the sphere size did influence the partitioning of the lipids: Fig. \ref{fig:figure3}b shows that only the smaller spheres are covered by LD domains, while the larger ones tend to have a composition similar to the surrounding substrate. Moreover, a few of the larger spheres have an irregular, bumpy, surface, and we observed POPC-rich domains on these dimples (indicated by yellow arrows in Fig. \ref{fig:figure3}b). This is a further manifestation of geometric pinning, in which the lipids going into the LD phase preferentially localize to regions of small radii of curvature.

We note that preference of the LD phase for locally spherical regions has also been observed by Subramaniam \emph{et al.} \cite{subramaniam2010particle}, where SLBs on patterned half-spherical PDMS caps were employed. In our case, however, we did not observe LD domains forming on the flat substrate for any composition, for waiting times of up to five days (see Fig.3 in the Supporting Information). This is an effect of the  nanoscopic substrate roughness which hinders the hydrodynamic flow of small domains, as explained by Goodchild {\em et al.} \cite{goodchild2019substrate}. The use of MICA substrates in place of glass can alleviate this effect \cite{goodchild2019substrate, Rinia2001}, and allow for the formation of LD domains of circular shape, see also Supporting Information, Fig.  3. While the substrate might hinder domain fusion, we have never observed hindered diffusion on the colloidal scaffolds, which instead enhance macroscopic phase-separation \cite{Madwar2015, Rinaldin2018}. Furthermore, non-zero mean curvature can promote phase separation, as the stability landscape of the mixture is shape-dependent \cite{sorre2009curvature,Rinaldin2018}. We further note that although we expected LD geometric pinning in the contact region between colloid and substrate, we have rarely observed localization of domains at such edges (see Supporting Information, Fig. 3). We hypothesize that this is due to partial detachment of the membrane from the colloid, allowing for a local decrease (in modulus) of the membrane curvature.

\begin{figure}
\includegraphics[width=\linewidth]{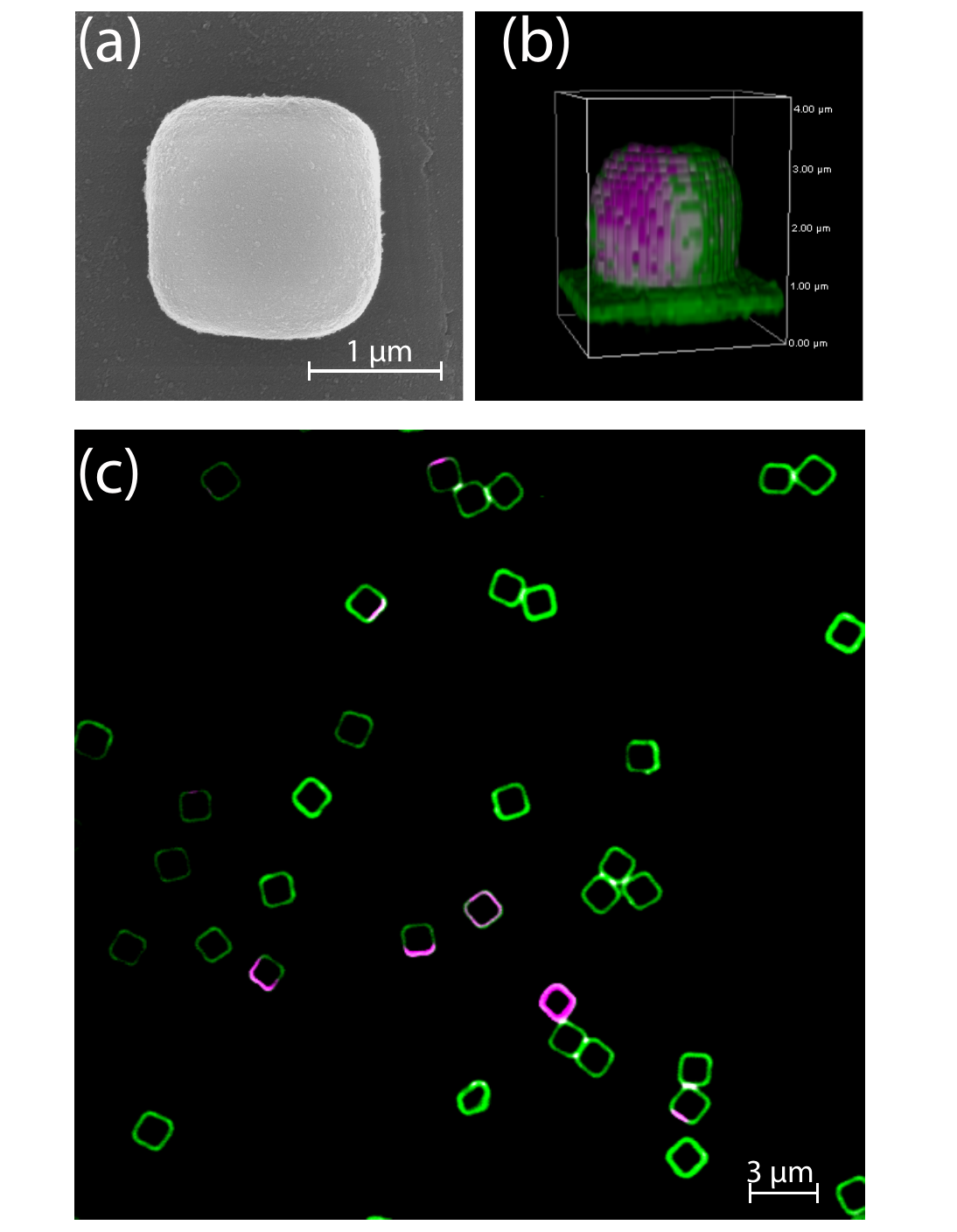}
\caption{
\textbf{(a)} Scanning electron microscopy image of a colloidal cubic silica shell. 
\textbf{(b)} 3D reconstruction of fluorescence microscopy image stack of a bilayer on a substrate patterned with cubic colloids. The LO phase is localized on the flat substrate and the colloid, the LD phase only on the colloid. 
\textbf{(c)} Fluorescence microscopy image of the equatorial plane of a lipid bilayer on a substrate  patterned with cubic colloids. }
\label{fig:figure4}
\end{figure} 

To investigate the effect of high mean curvature and negative Gaussian curvature on the Gibbs triangle, we have repeated the same analysis for substrates patterned with dumbbell-shaped particles. 
The dumbbells were obtained by inducing protrusions on PS-TPM spheres, which were subsequently coated with silica (as in Fig. \ref{fig:setup}b) for SLB formation. The colloids were calcinated at \SI{450}{\celsius}, such that they consist of a single shell of silica. In some cases, this shell can break and indent the surface. We determined the ternary phase diagram for substrates patterned both with symmetric and asymmetric dumbbell particles and plot our results in a Gibbs phase triangle in Fig. \ref{fig:figure3}c. Similarly to the case of substrates patterned with spherical colloids, we first distinguish between mixed and demixed configurations. The former state is indicated by red squares in the diagram. However, different with respect to spheres (Fig. \ref{fig:figure3}a), we can further differentiate phase-separated states depending on whether LD domains wrap the entire colloid (blue circles) or whether they localize only along the neck region of the dumbbell (yellow triangles). We label these two phase-separated states respectively as type I and II. Thus, spheres exhibit only type I demixed configurations.

Comparing Fig. \ref{fig:figure3}c with Fig. \ref{fig:figure3}a, we observe that the binodal line, separating mixed form demixed configurations, is rather insensitive, within the resolution of our sampling, to the geometry of the colloids. In the phase-separated region of the phase diagram of the dumbbell particles, it appears that cholesterol is the main discriminant between type I and type II states. Since cholesterol has also been previously found to influence domain size distributions \cite{pathak2011measurement}, we conjecture that Fig. \ref{fig:figure3}c shows a similar effect. Furthermore, we observe that symmetric and asymmetric dumbbells exhibit a very similar behaviour that is dominated by the highly curved neck region, as opposed to what we observed in SLVs \cite{Rinaldin2018}. Therefore, we conclude that for open membranes, the relative curvature difference between the two dumbbell lobes is not as important as for closed SLVs. Finally, in line with the theoretical predictions,\cite{fonda2019thermodynamic,fonda2018interface} we have never found any signs of antimixing on any of the membrane shapes in contact with a reservoir, emphasizing the significant difference induced by opening up the membrane. 

Dumbbells and spheres were not the only shapes that were considered: in Fig. \ref{fig:figure4} we show a representative image of a substrate patterned  with cuboidal particles. Here, the inhomogeneous curvature of the colloids induces lipid segregation as observed similar to the case of spheres and dumbbells. However, we were not able to see any striking geometric pinning. We hypothesise that this is due to the small curvature differences of the surfaces of the cube which hinder a pinning of the softer phase on the whole surface of the colloid. This result is similar to what we previously observed in SLVs \cite{Rinaldin2018} in which there was no significant correlation between the position of the LD domains and the curvature of the cubic supports. 
 
\begin{figure*}[ht]
\includegraphics[width=0.8\linewidth]{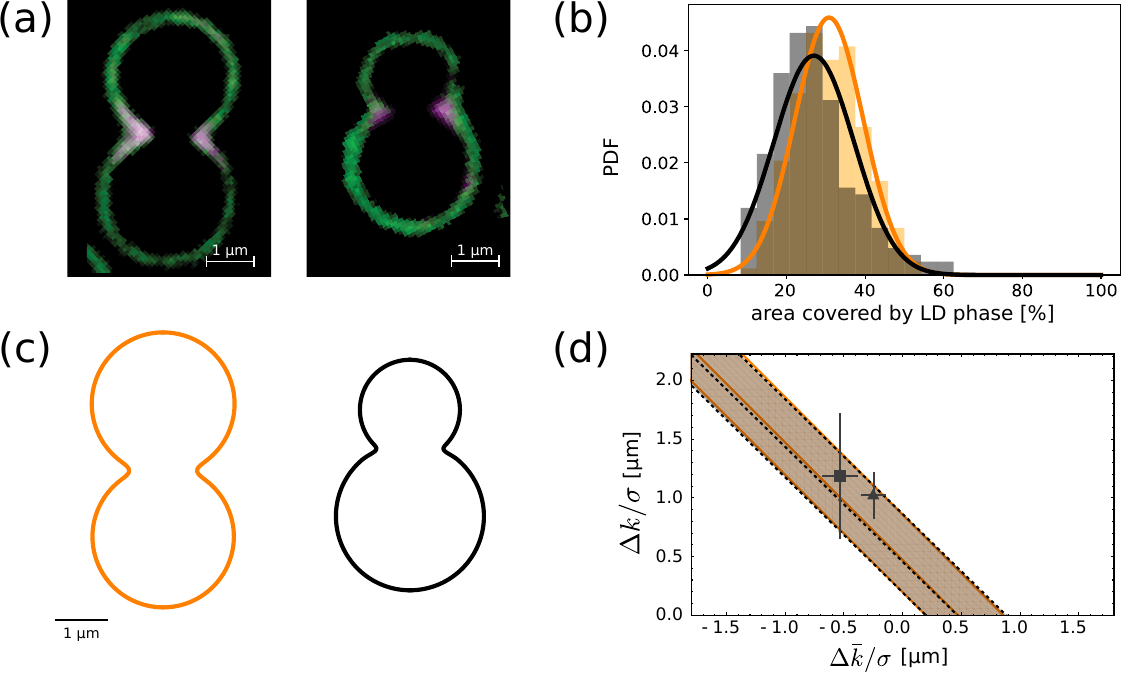}
\caption{
\textbf{(a)} Fluorescence microscopy image of the equatorial profile of a membrane on an asymmetric and symmetric dumbbell-shaped colloid.
\textbf{(b)} LD area fraction histograms for symmetric (orange) and asymmetric (black) dumbbell-shaped SLVs. The histograms are fitted with Gaussian distributions, with means and variance respectively at $31 \pm 9\%$ and $27 \pm 10 \%$. 
\textbf{(c)} Reconstructed profiles for the typical symmetric and asymmetric dumbbell geometries. The shape of the necks is obtained by a fourth-order smooth polynomial interpolation between two spherical caps (see also \cite{fonda2019thermodynamic}). The length scale is chose so to match experimental images. 
\textbf{(d)} In the parameter space spaced by the curvature moduli differences divided by the line tension, we show the allowed region implied by compatibility - at one standard deviation - from the neck domain size distributions of (b). Both shapes point at essentially the same strip-like region (delimited by continuous orange and by dashed black lines), which is compatible with previous measurements done on GUVs by \cite{baumgart2005membrane} (grey square with error bars) and \cite{semrau2008accurate} (grey triangle). 
}
\label{fig:figure5} 
\end{figure*} 

\subsection{Quantification of geometric pinning in dumbbell-shaped colloids}

The image in Fig. \ref{fig:figure3}d suggests that the size and localisation of the LD domains along neck regions is rather constant. To further investigate this, we measured the area fraction occupied by each domain on a given colloid. We collected data for 200 distinct colloidal particles of each kind taken from a SLB with composition of $50\%$ BSM, $25\%$ POPC and $25\%$ Cholesterol, see Fig. \ref{fig:figure3}c. The area fraction was computed from measurements of the normalised fluorescence intensity of the mid-plane profile, such as the ones shown in Fig. \ref{fig:figure5}a, under the assumption that the shape of both domains is axisymmetric. Although the presence of the substrate breaks rotational invariance of the membrane, this is still a reasonable approximation for the \textit{relative} area fractions. 
Fig. \ref{fig:figure5}b shows the size distributions for symmetric (orange) and asymmetric (black) dumbbells. A Gaussian fit of the histograms gives mean area fractions of  $31 \pm 9 \%$ and $27 \pm 10 \%$, respectively. Thus, there is no significant difference in the area fractions of LD domains between symmetric and asymmetric colloids in the presence of a lipid reservoir. 
 
To test if the total lipid composition affects the typical area fraction occupied by the LD domains, we furthermore measured the LD area fraction at the neck of the dumbbell shaped particles for three different SLB compositions of BSM:POPC:Chol, namely 50:25:25, 40:20:40, and 20:50:30. We analysed 50 colloidal symmetric dumbbell-shaped particles and found that the area fractions of the pinned domains are  $31 \pm 9 \%$,  $26 \pm 6 \%$, and  $26 \pm 4 \%$, respectively. We thus conclude that the global composition of the membrane also does not significantly affect the size of the LD domains.

The data presented above can be used to estimate the difference in the bending moduli of the LO and LD phases. For this purpose, we model the membrane using the J\"ulicher-Lipowksy free energy for multi-component bilayers \cite{Julicher1993}, which describes the membrane as a two-dimensional multi-phase stationary fluid, whose free-energy depends upon the local mean curvature $H$ and Gaussian curvature $K$, namely:
\begin{equation}
F = \sum_{i={\rm LD,\,LO}} \int_{\Sigma_{i}} {\rm d}A\, (k_{i}H^{2}+\bar{k}_{i}K)+\sigma \oint_{\Gamma} {\rm d}s\,, 	
\end{equation}
where $\Sigma_{i}$ denotes the, possibly disconnected, portion of the membrane occupied by the LO and LD phases and $\Gamma$ their interface. The constants $k_{i}$, $\bar{k}_{i}$ and $\sigma$ embody, respectively, the stiffness of the lipid phases with respect to bending and Gaussian splay, while $\sigma$ indicates the tension of the LO/LD interface.
We neglect contributions to the free energy due to spontaneous curvature by assuming two leaflets of the bilayer to be identical. Note that often in the literature there is an extra factor of 2 in the definition of $k_i$. Within this framework, the configuration of the domains is determined by only two parameters: $\Delta k / \sigma$ and $\Delta \bar{k} / \sigma$, where $\Delta k = k_{\mathrm{LO}} - k_{\mathrm{LD}}$, $\Delta \bar{k} = \bar{k}_{\mathrm{LO}} - \bar{k}_{\mathrm{LD}}$. 
In \cite{fonda2018interface} we proved that a necessary condition for an LO/LD interface in an open membrane to be at equilibrium is that
\begin{equation}
\kappa_g = \frac{\Delta k}{\sigma} H^2 + \frac{\Delta \bar{k}}{\sigma} K \,, 
\label{eq:interface}
\end{equation}
where $\kappa_g$ is the interface's geodesic curvature. Applying Eq. \eqref{eq:interface} to the case of an axisymmetric membrane is a rather straightforward task as described in the Supporting Information. The final equation can be solved numerically for a given profile shape, such as the ones of Fig. \ref{fig:figure5}c. These profiles are constructed by joining two spherical caps via a smooth polynomial neck. Their absolute size is fixed by the experimental system (Fig. \ref{fig:figure5}a) leading to areas of \SI{44.6}{\micro\meter^2} (symmetric dumbbell, orange) and \SI{28.5}{\micro\meter^2} (asymmetric dumbbell, black) and with minimal neck radius of \SI{650}{\nano\meter} and \SI{580}{\nano\meter}, respectively. 	

\begin{figure*}[ht]
\includegraphics[width=1\linewidth]{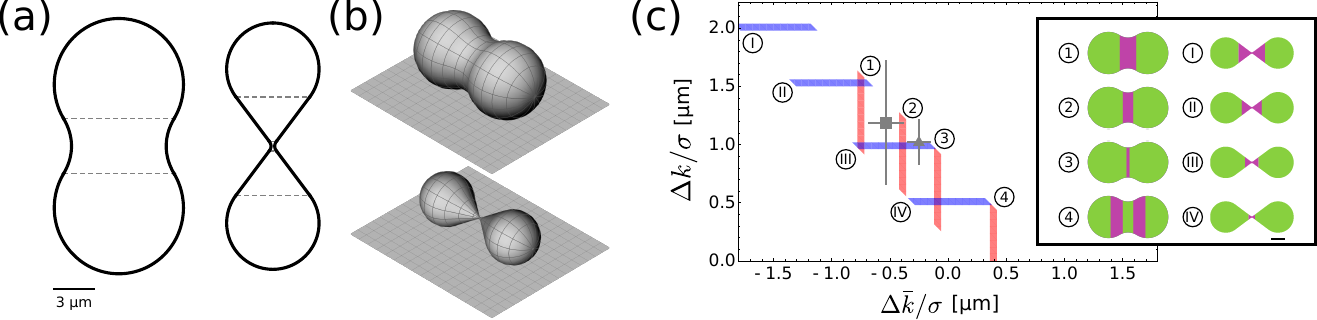}
\caption{
\textbf{(a)} Proposed shapes for dumbbells with catenoidal (left) or conical (right) necks, to be 3D printed onto a substrate. Note that the scale bar is \SI{3}{\micro\metre}, so that these structures are much larger than the ones in Figs. \ref{fig:figure5}a-c. The horizontal dashed lines show the smooth gluing points between the neck and the spherical caps.
\textbf{(b)} 3D rendering of the profiles in (a).
\textbf{(c)} As in Fig. \ref{fig:figure5}D, we show the overlap between our previous estimates for the parameters and an imaginary measurement of an interface on the catenoidal dumbbell (vertical red regions) or on the conical dumbbell (horizontal blue regions. The two gray data points refers to previous measurements \cite{baumgart2005membrane,semrau2008accurate}. The inset on the right shows the actual position of the interface for the two geometries and four different values.}
\label{fig:figure6} 
\end{figure*} 

By comparing the area fraction histograms of Fig. \ref{fig:figure5}b with the solutions of Eq. \eqref{eq:interface}, it is possible to estimate the two parameters $\Delta k/ \sigma$ and $\Delta \bar{k}/ \sigma$, both having units of length. The outcome of this comparison is summarised in Fig. \ref{fig:figure5}d. In the plane spanned by the two parameters, the shaded region corresponds to the values that are compatible with experimental observations within one standard deviation. Interestingly, the two geometries carry essentially the same information as the two shaded regions are completely overlapping. Furthermore, the region has the shape of a straight strip with slope $-1$, demonstrating how these geometries are only sensitive to the combination $(\Delta k+ \Delta \bar{k})/ \sigma$ rather than the two parameters independently. This property ultimately originates from the fact that the observed interfaces lie on the spherical portion of the colloid. Since a sphere of radius $R$ has $H^2=K=1/R^2$ everywhere, Eq. \eqref{eq:interface} depends only on the sum of the two parameters, therefore the allowed regions must lie on a strip \cite{fonda2018interface}. Our findings are fully compatible with measurements on multicomponent GUVs available in the literature \cite{baumgart2005membrane,semrau2008accurate}. We find that $(\Delta k+ \Delta \bar{k})/ \sigma$ is equal to $0.47 _{-0.28}^{+0.38}$ \SI{}{\micro\meter} for us, while other groups had previously obtained \SI{0.77\pm 0.21}{\micro\meter} \cite{semrau2008accurate} and \SI{0.65\pm 0.54}{\micro\meter} \cite{baumgart2005membrane}. 

\subsection{Forecast of experiments on catenoidal and conical necks}

The umbilical~\footnote{A point on a surface is umbilic whenever $H^2=K$.} nature of spherical caps prevents a separate quantification of the two parameters. Such a measurement could be achieved only for interfaces lying on non-spherical, curved parts of the patterned substrate. Therefore, we here propose shapes and sizes which could provide a more precise and independent estimation of   $\Delta k/ \sigma$ and $\Delta \bar{k}/ \sigma$. One approach would be to have an interface on a portion of a surface where either of the two terms in the right hand side of Eq. \eqref{eq:interface} vanishes, i.e. when either $H=0$ or $K=0$. In the former case, the surface must be minimal, while in the latter it must be developable. Surfaces with a neck that have these properties and retain the axisymmetric structure are uniquely determined: they must be a catenoid or a circular cone, respectively. Since none of these are compact surfaces, we can smoothly close them by gluing them to spherical end caps. The resulting geometry for the patterned feature is either a catenoidal or a conical dumbbell, depicted in Fig. \ref{fig:figure6}a and b. 

The geometry of these surfaces allows us to solve Eq. \eqref{eq:interface} analytically under the assumption that the interface is parallel, see the Supporting Information. Both shapes depend on just one free parameter, respectively the catenoid radius and the cone slope. We can choose these parameters in order to maximize the chances of measuring $\Delta k/ \sigma$ and $\Delta \bar{k}/ \sigma$ compatible with known values. The results are the shapes shown in Fig \ref{fig:figure6}a. We report how different values of the moduli differences would produce different pinning patterns in Fig. \ref{fig:figure6}c. Specifically, we forecast that a catenoidal dumbbell of length \SI{18.5}{\micro\metre} and neck minimal radius of \SI{3.5}{\micro\metre} would make it easiest to discriminate between values of $\Delta \bar{k}/ \sigma$ of similar magnitude to known ones. Similarly, we find that a conical dumbbell of length \SI{17.8}{\micro\metre} and with slope $3/4$ is optimal for measuring the difference in bending moduli. Substrates with these geometrical features can in principle be obtained by  using a combination of 3D direct laser writing and replica-molding with PDMS \cite{rinaldin2019on}. 

\section{Conclusions}

In our previous work \cite{Rinaldin2018}, we have shown that closed lipid membranes supported on colloidal particles (so called scaffolded lipid vesicles, or, SLVs) feature domains whose position and composition are strongly affected by curvature and the overall membrane composition. In spherical SLVs of constant curvature, we observed the coexistence of two domains as result of minimisation of line tension. In dumbbell-shaped SLVs we have previously identified the presence of geometric pinning and antimixing, i.e. a state in which lipids are mixed and yet organized in domains with strikingly different compositions. As we have shown in theoretical analyses \cite{fonda2018interface,fonda2019thermodynamic}, these states are a consequence of the fact that the membrane is at the same time closed and scaffolded. 

In this paper, we test this hypothesis experimentally by relieving the constraint of membrane closeness by connecting SLVs to a reservoir of lipids. To this end, we deposited colloidal particles on a flat substrate and covered both particles and substrate with a continuous multicompontent lipid bilayer. In this setup, the lipid membrane region on the colloid is open in the sense that it can continuously exchange lipids with the membrane on the flat substrate. The latter can be considered as a reservoir of lipids for the bilayer on the colloids. Using spherical particles of different radii as well as symmetric and asymmetric dumbbell-shaped particles, we showed that the equilibrium landscape changes dramatically when the membrane is opened up. Instead of the composition dependency which was found for SLVs, we here observed a highly regular localisation of the softer LD phase in regions of higher mean curvature and negative Gaussian curvature for a range of compositions. The lipid exchange enhanced the geometric pinning of the LD phase to small spherical and dumbbell-shaped membrane regions as well as on the neck of the dumbbell-shaped membranes. For spheres, our results are in agreement with previous observations on substrates with half-spherical asperities \cite{subramaniam2010particle}. However, we here have also reported how this behaviour depends on the total lipid composition of the supported lipid bilayer in Gibbs phase triangles for both spheres and dumbbells. 

The regular pinning of the lipid domains on the neck of the dumbbells furthermore allowed us to measure the size of the disordered domains with higher precision than previously \cite{Rinaldin2018}. By combining the experimental results with numerical simulations, we were able to obtain a precise estimation of the combination of the elastic material parameters $(\Delta k+ \Delta \bar{k})/ \sigma$, which was found to be $0.47 _{-0.28}^{+0.38}$ \SI{}{\micro\meter}. Interestingly, our result is compatible with the two other measurements available in the literature obtained from free-standing GUVs \cite{baumgart2005membrane,semrau2008accurate}. This agreement indicates that our measurements are related to true membrane properties rather than being linked to the specific experimental set-up. In particular, the role of the adhesion energy to the scaffold ,which might be relevant in our experiments, seems to be negligible.

Finally, we have proposed membrane shapes that would allow a more precise and independent measurement of these material parameters by disentangling mean and Gaussian curvatures,  i.e. dumbbell-shaped membranes with either catenoidal or conical necks. We have furthermore determined the size and curvature that are necessary to observe disordered domains and enable the experimental determination of material parameters. These geometries are not just hypothetical, since substrates and hence membranes of designed 3D shape can in principle be obtained by a combination of micro-printing and replica-molding \cite{rinaldin2019on}. Our work offers a collection of experimental and theoretical tools as well as new insights into the effect of lipid exchange in multicomponent bilayers.

\paragraph*{Acknowledgements} 
This work was supported by the Netherlands Organisation for Scientific Research (NWO/OCW), as part of the Frontiers of Nanoscience program (MR), the VIDI scheme (LG,PF) and by the European Research Council (ERC) under the EU Horizon 2020 research and innovation program (DJK, grant agreement no. 758383). We thank Vera Meester and Rachel Doherty for help with particle synthesis and electron microscopy imaging.

\bibliography{references}

\newpage
\begin{figure*}[ht]
\includegraphics[width=0.5\linewidth]{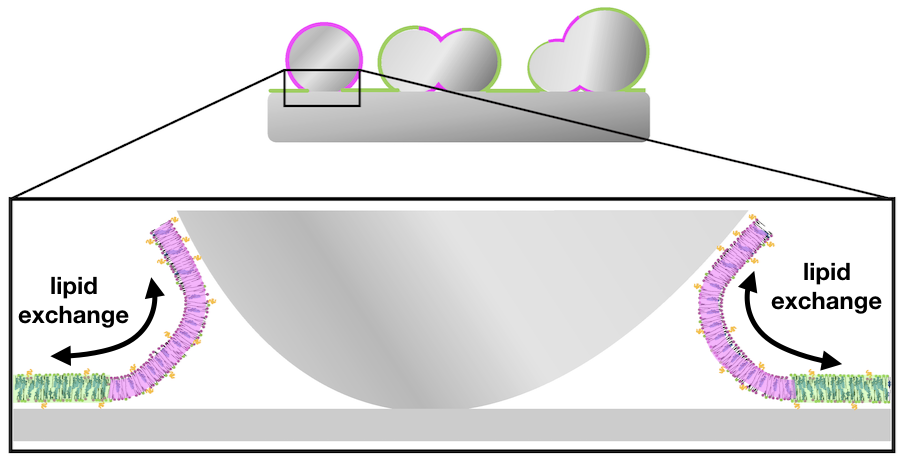}
\caption{Table of Content entry: "Local lipid exchange is crucial in determining the phase behaviour of multicomponent membrane-coated colloidal particles"}
\end{figure*} 

\end{document}